\documentclass[aps,prd,twocolumn,superscriptaddress,nofootinbib,10pt]{revtex4-1}   

\usepackage[utf8]{inputenc}
\usepackage{amssymb}
\usepackage{amsmath}
\usepackage{amsfonts}
\usepackage{graphicx}
\usepackage{color}
\usepackage{xspace}
\usepackage{comment}
\usepackage{hyperref}
\usepackage[normalem]{ulem}

\usepackage[section]{placeins}
\usepackage{afterpage}

\usepackage{float}
\usepackage{slashed}
\usepackage{appendix}

\usepackage{cancel}

\usepackage{multirow,rotating}
\usepackage[dvipsnames]{xcolor}
\usepackage{orcidlink}   

\begin{document}
	
      \title{Probing the dark axion portal via $J/\psi$ decays at BESIII and STCF}

	   \author{Zeren Simon Wang\,\orcidlink{0000-0002-1483-6314}}
      \affiliation{School of Physics, Hefei University of Technology, Hefei 230601, People’s Republic of China}

      \author{Dazhuang He\,\orcidlink{0009-0001-3444-8378}}
      \email{dzhe1998@gmail.com}
      \affiliation{College of Physics and Electronic Engineering, Heze University,
No.2269 University Road, Heze, Shandong, 274015, P.R.China}

      \author{Yu Zhang\,\orcidlink{0000-0001-9415-8252}}
      \email{dayu@hfut.edu.cn}
      \affiliation{School of Physics, Hefei University of Technology, Hefei 230601, People’s Republic of China}

	\begin{abstract}      
        Large numbers of $J/\psi$ mesons can be resonantly produced at BESIII and STCF at the center-of-mass energy $\sqrt{s}=3.097$~GeV. Such $J/\psi$ mesons may undergo rare decays into an axionlike particle (ALP) $a$ and a dark photon $\gamma'$ through the dark axion portal. We investigate the exclusion reach of the existing BESIII dataset and the projected sensitivity of STCF, focusing on the mono-photon signature. We perform Monte Carlo simulations of the signal and estimate both the irreducible $J/\psi\to\gamma\nu\bar{\nu}$ background and the continuum and resonant three-photon reducible backgrounds. Detector-induced migration of three-photon events into the mono-photon signal region is parameterized by the event-level leakage probability $\epsilon_{3\gamma}^{\rm leak}=0,10^{-6},10^{-4},10^{-3}$. In the zero- and sufficiently small-leakage benchmarks, the existing BESIII dataset could probe previously unexplored parameter space, while STCF can further improve the coupling reach. Increasing leakage probabilities progressively weaken these projections, highlighting the importance of detector-level three-photon rejection in realizing the projected reach.
 	\end{abstract}



	\maketitle
    \noindent

\section{Introduction}\label{sec:intro}

In recent years models of portal physics are under scrutiny, as to whether they serve as a bridge between the Standard-Model (SM), visible sector and a dark, invisible sector.
Traditionally, such models include the fermion portal with heavy neutral leptons~\cite{Shrock:1980vy,Shrock:1980ct,Shrock:1981wq}, the scalar portal with a new scalar particle~\cite{OConnell:2006rsp,Wells:2008xg,Bird:2004ts,Pospelov:2007mp,Krnjaic:2015mbs,Boiarska:2019jym}, the vector portal with a dark photon $\gamma'$~\cite{Okun:1982xi,Galison:1983pa,Holdom:1985ag,Boehm:2003hm,Pospelov:2008zw,Curtin:2014cca}, and the pseudoscalar portal with an axion or axionlike particle (ALP) $a$~\cite{Peccei:1977hh,Peccei:1977ur,Witten:1984dg,Conlon:2006tq,Arkani-Hamed:2006emk,Arvanitaki:2009fg,Cicoli:2012sz}.
The predicted new particles listed above couple to the SM particles either directly or via mixing with them.
An additional portal, called the dark axion portal (DAP), was introduced more recently in Ref.~\cite{Kaneta:2016wvf}.
As an independent portal, the DAP provides a novel link between the dark photon and the ALP, featured with the coupling $G_{a\gamma\gamma'}$, instead of being a naive combination of the vector and pseudoscalar portals.
The DAP can be constructed from a triangle anomaly, in the underlying UV-complete models such as the dark KSVZ axion model discussed in Ref.~\cite{Kaneta:2016wvf}.

The DAP scenario is well motivated, for solving critical issues in particle physics and cosmology, including the strong CP problem~\cite{Kaneta:2016wvf,Kaneta:2017wfh,Broadberry:2024pkv} and the hierarchy problem with a cosmological relaxation solution~\cite{Graham:2015cka,Choi:2016kke,Domcke:2021yuz}, as well as providing a dark-matter candidate~\cite{Kaneta:2016wvf,deNiverville:2019xsx,Zhevlakov:2022vio,Gninenko:2026mgn}.
For further past studies on the dark axion portal, we refer to Refs.~\cite{deNiverville:2018hrc,Kalashev:2018bra,Deniverville:2020rbv,Hook:2021ous,Hook:2023smg,Hong:2023fcy,Jodlowski:2023yne,Jodlowski:2024ayf,Ness:2025klj,Shen:2026kzy}.
Such studies cover contexts of cosmological observations, colliders including $B$-factories, beam-dump experiments, astrophysics, and neutrino experiments.

In this work, we investigate the sensitivity of the existing BEijing Spectrometer III (BESIII) experiment at the Beijing Electron Positron Collider II (BEPC~II)~\cite{BESIII:2009fln} in Beijing, China, as well as that of the proposed Super Tau--Charm Facility (STCF)~\cite{Achasov:2023gey,Ai:2025xop} in Hefei, China, to the dark axion portal.
BESIII and STCF are intensity-frontier electron-positron collider experiments, scanning the center-of-mass (COM) energy over about 2--5~GeV and 2--7~GeV, respectively, in the tau--charm regime.
In particular, both experiments can operate at the COM energy of $3.097$~GeV  on the $J/\psi(1S)$ resonance,  yielding $\mathcal{O}(10^{10})$ and $\mathcal{O}(10^{12})$ $J/\psi$ mesons , respectively.
The $J/\psi$ mesons can undergo two-body decays via an off-shell SM photon $\gamma$ into a dark photon and an ALP in the DAP scenario, i.e., $J/\psi\to a\gamma'$, as studied in the context of the LHC far detectors~\cite{Jodlowski:2023yne} and the beam-dump experiments NA64$e$, NA64$\mu$, LDMX, and M$^3$~\cite{Zhevlakov:2022vio,Gninenko:2026mgn}.  

We focus on the phenomenology with the single coupling $G_{a\gamma\gamma'}$ in this work.
With this coupling present, an additional coupling $G_{a Z \gamma'}$ arises from the SM gauge invariance, with the relation $|G_{a Z \gamma'}/G_{a\gamma\gamma'}|=\tan\theta_W$ where $\theta_W$ is the weak mixing angle~\cite{Jodlowski:2024ayf}.
Since we consider BESIII and STCF  on the $J/\psi(1S)$ resonance, the $Z$-boson contributions to the signal process are negligible, and we will therefore ignore the coupling $G_{a Z \gamma'}$ in our phenomenological analysis.

The sensitivity projections presented in this work are based on simulations and analytic estimates of the signal and background contributions.
Although we make every effort to model the relevant BESIII detector response and the anticipated STCF performance, our results remain phenomenological estimates.
In particular, the background estimates and the resulting sensitivity projections may be optimistic.
They will ultimately require validation and refinement by the BESIII and future STCF collaborations using full detector simulations and data-driven studies based on control samples.

This paper is organized as follows.
In Sec.~\ref{sec:theory} we present the model setup, computing the decay rates of the $J/\psi$ meson, dark photon, and ALP.
Then in Sec.~\ref{sec:experiments_simulation} we introduce the BESIII and STCF experiments, detail our simulation procedures, as well as elaborate on background estimates.
We show the numerical results in Sec.~\ref{sec:results} and conclude the paper in Sec.~\ref{sec:conclusions}.
Additionally, Appendix~\ref{appendix:Jpsi_decay} provides the analytic differential decay rates for the signal and irreducible background, while Appendix~\ref{appendix:threegamma_background} details our benchmark estimate of the reducible three-photon backgrounds.

\section{Theoretical setup}\label{sec:theory} 

We focus on the following interaction Lagrangian with a single term proportional to the coupling $G_{a\gamma\gamma'}$~\cite{Kaneta:2016wvf,Ejlli:2016asd}
\begin{eqnarray}
    \mathcal{L}_{\text{DAP}}\supset\frac{G_{a\gamma\gamma'}}{2} \, a \, F_{\mu\nu} \, \tilde{Z}^{\prime\mu\nu}, \label{eq:Lag_DAP}
\end{eqnarray}
where $G_{a\gamma\gamma'}$ has mass dimension $[-1]$, and $F_{\mu\nu}$ is the electromagnetic field-strength tensor and $Z^{\prime\mu\nu}$ is the field-strength tensor of the dark U(1)$_D$ gauge field.
$\tilde{Z}^{\prime\mu\nu}$ is the dual of $Z^{\prime\mu\nu}$.

As discussed above, the interaction in Eq.~\eqref{eq:Lag_DAP} may induce $J/\psi$ decays into a dark photon and an ALP, and we compute the corresponding decay width to be
\begin{eqnarray}
\Gamma(J/\psi\to a\gamma')=\frac{e^2Q_c^2f_{J/\psi}^2  G_{a\gamma\gamma'}^2}{96\pi\,m_{J/\psi}^5}\, \lambda^{3/2} (m_{J/\psi}^2,m_a^2,m_{\gamma'}^2),\label{eq:GammaJpsi2agammaPrime}
\end{eqnarray}
where $e$ denotes the electric charge coupling, $Q_c=2/3$ is the electric charge in units of $e$, $f_{J/\psi}$ and $m_{J/\psi}$ are the decay constant and mass of $J/\psi$, respectively, and $\lambda(x,y,z)=x^2+y^2+z^2-2xy-2xz-2yz$ labels the K\"all\'en function.
$m_a$ and $m_{\gamma'}$ are the masses of the ALP and the dark photon, respectively.

We notice on the other hand that the decay width of the SM process $J/\psi\to e^+e^-$~\cite{QuarkoniumWorkingGroup:2004kpm} is,
\begin{eqnarray}
\Gamma(J/\psi\to e^+e^-)=\frac{4\pi\alpha^2Q_c^2f_{J/\psi}^2}{3m_{J/\psi}},
\end{eqnarray}
where $\alpha=e^2/4\pi$ is the fine structure constant.
We thus express $\Gamma(J/\psi\to a\gamma')$ in terms of $\Gamma(J/\psi\to e^+e^-)$,
\begin{eqnarray}
\Gamma(J/\psi \to a\gamma')&=&\frac{G_{a\gamma\gamma'}^2 \, \Gamma(J/\psi\to e^+e^-)}{32\,\pi\,\alpha\,m_{J/\psi}^4}\nonumber\\
&&\lambda^{3/2}(m_{J/\psi}^2,m_a^2,m_{\gamma'}^2),\label{eq:GammaJpsi2agammaPrime_replaced}
\end{eqnarray}
where $\Gamma(J/\psi\to e^+e^-)=\Gamma_{J/\psi}\cdot\text{BR}(J/\psi\to e^+e^-)=5.53$~keV with the measured $J/\psi$ total decay width $\Gamma_{J/\psi}=92.6$~keV and $\text{BR}(J/\psi\to e^+ e^-)=5.971\%$~\cite{ParticleDataGroup:2024cfk}.

\begin{figure}[t]
    \centering
    \includegraphics[width=0.99\linewidth]{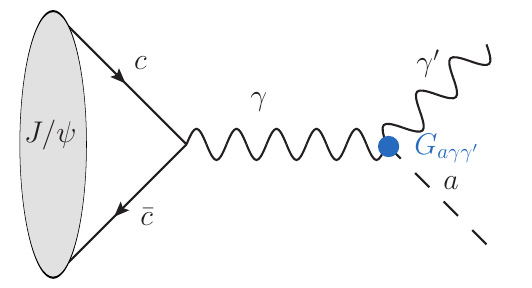}
    \caption{The Feynman diagram depicting the $J/\psi$ decay into a dark photon and an ALP via an $s$-channel SM photon.}
    \label{fig:feynman}
\end{figure}

In Fig.~\ref{fig:feynman} we show the Feynman diagram of the decay $J/\psi\to a\gamma'$ via an off-shell SM photon.
The vertex associated with the coupling $G_{a\gamma\gamma'}$ is labeled with a blue blob.

\begin{figure}[t]
    \centering
    \includegraphics[width=0.99\linewidth]{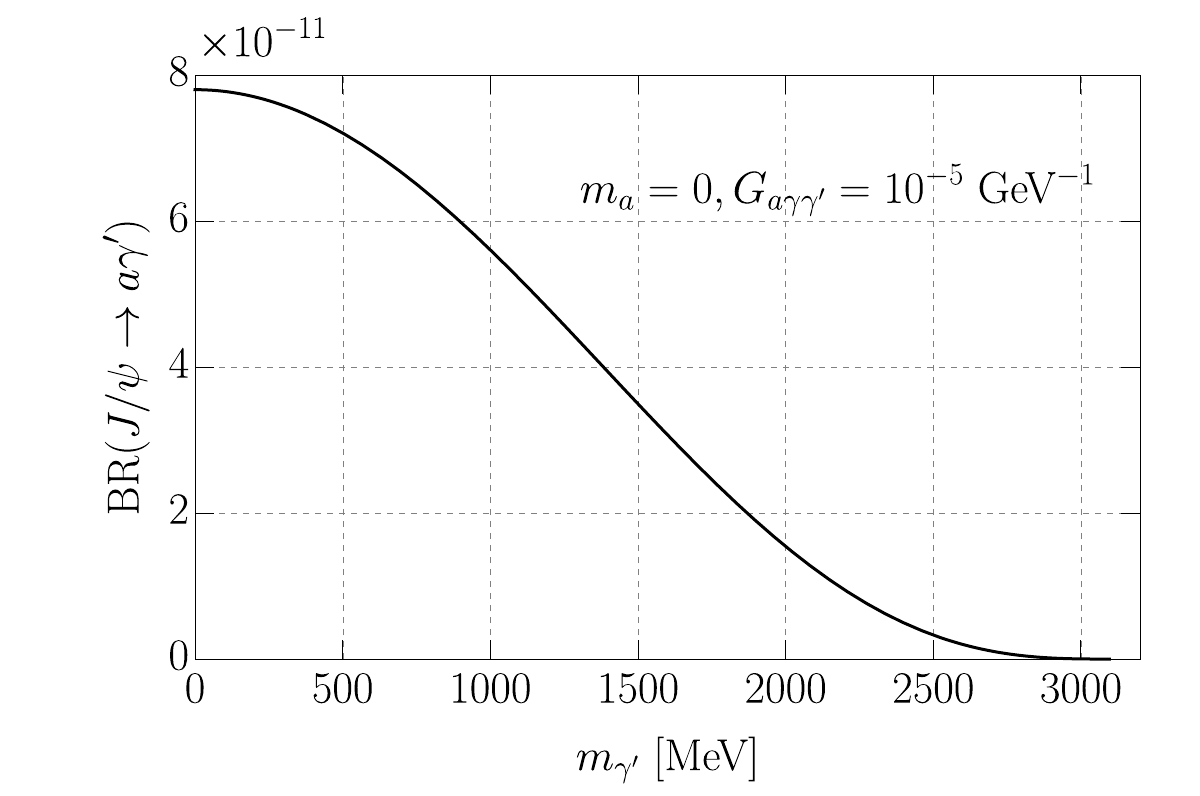}
    \caption{The decay branching ratio of $J/\psi\to a\gamma'$ as a function of $m_{\gamma'}$ for $m_a=0$ and $G_{a\gamma\gamma'}=10^{-5}$~GeV$^{-1}$. }
    \label{fig:BR_Jpsi}
\end{figure}

Further, we display in Fig.~\ref{fig:BR_Jpsi} a plot of BR$(J/\psi\to a\gamma')=\Gamma(J/\psi\to a\gamma')/\Gamma_{J/\psi}$ as a function of $m_{\gamma'}$, for $m_a=0$ and $G_{a\gamma\gamma'}=10^{-5}$~GeV$^{-1}$.
For this benchmark,  the branching ratio is of order $10^{-11}$.
$\Gamma(J/\psi \to a\gamma')$ (see Eq.~\eqref{eq:GammaJpsi2agammaPrime_replaced}) is symmetric under $m_a\leftrightarrow m_{\gamma'}$ and scales as $G_{a\gamma\gamma'}^2$.

The leading two-body decays induced by the $G_{a\gamma\gamma'}$ coupling and their corresponding partial widths are:
\begin{eqnarray}
    \Gamma(\gamma'\to a \gamma) &=&\frac{G^2_{a\gamma\gamma'}}{96\pi} m^3_{\gamma'}\Big( 1- \frac{m_a^2}{m^2_{\gamma'}}\Big)^3,\label{eq:gammaPrimedecayrate}\\
    \Gamma(a\to\gamma' \gamma) &=&\frac{G^2_{a\gamma\gamma'}}{32\pi}m_a^3\Big(  1-\frac{m^2_{\gamma'}}{m_a^2}  \Big)^3.\label{eq:ALPdecayrate}
\end{eqnarray}
The two modes are kinematically allowed for $m_{\gamma'}>m_a$ and $m_a>m_{\gamma'}$, respectively.

We work in a simplified benchmark in which $G_{a\gamma\gamma'}$ is the only new-physics coupling relevant at the $J/\psi$ energy scale.
In particular, dark-photon kinetic mixing and additional ALP couplings to SM particles are assumed to be negligible.
If appreciable, such interactions would open decays such as $\gamma'\to \ell^+\ell^-$, $a\to\ell^+\ell^-$, and $a\to\gamma\gamma$, and could lead to experimentally cleaner final states containing charged-lepton pairs, possibly accompanied by a photon and/or missing energy.
These signatures depend on additional independent couplings and lie outside the scope of our single-coupling mono-photon analysis.

Even within this single-portal-coupling benchmark, the photon in the
radiative transition can be off-shell, giving rise to the Dalitz-type three-body decays
$\gamma'\to a\gamma^*\to a f\bar f$ and $a\to\gamma'\gamma^*\to\gamma' f\bar f$, when kinematically allowed.
These modes are suppressed by the additional electromagnetic coupling and the three-body phase space, and typically contribute only at the percent level to the total decay width~\cite{Jodlowski:2023yne,Jodlowski:2024ayf}.
We therefore neglect their effects on the lifetimes and mono-photon sensitivities considered here.

\begin{figure}[t]
    \centering
    \includegraphics[width=0.99\linewidth]{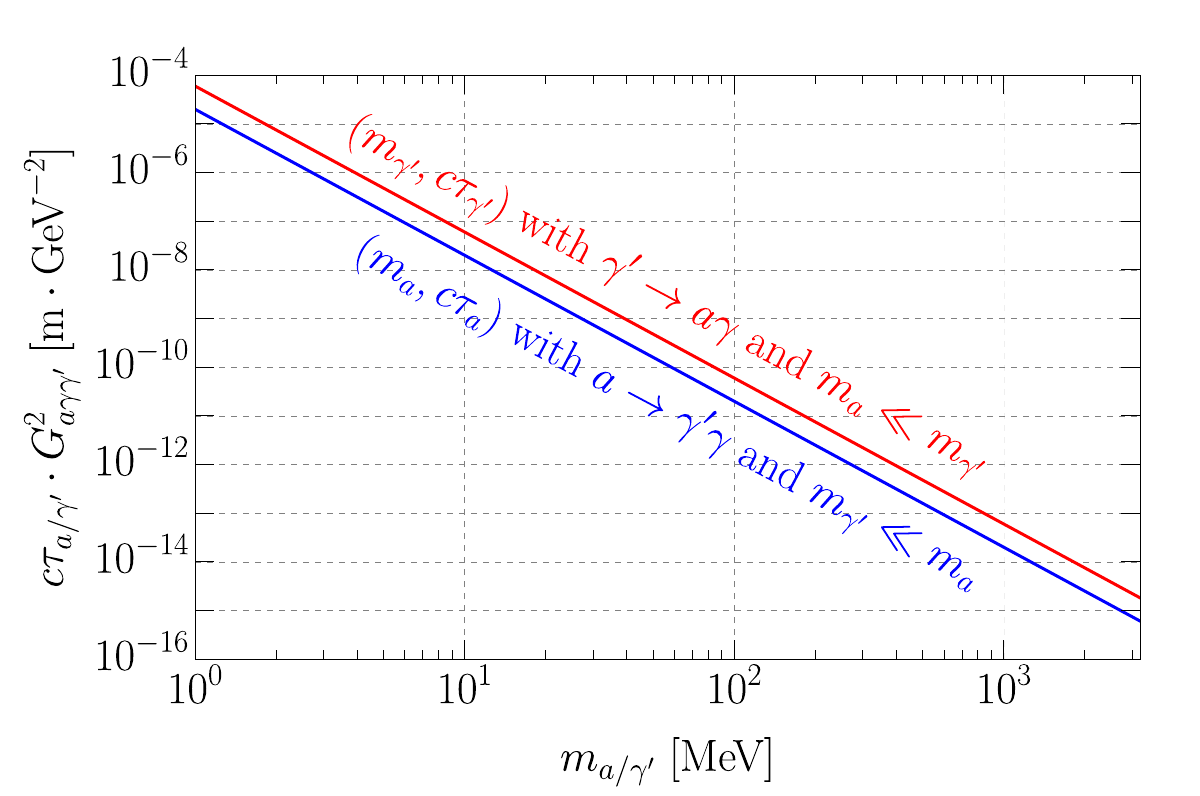}
    \caption{The proper decay length $c\tau_{a/\gamma'}$ of the heavier new particle ($\gamma'$ or $a$) scaled with $G^2_{a\gamma\gamma'}$ as a function of its mass, with the other particle being much lighter or massless.}
    \label{fig:ctau}
\end{figure}

The proper decay lengths $c\tau_a$ and $c\tau_{\gamma'}$  are computed  as $c\hbar/\Gamma_{a}$ and $c\hbar/\Gamma_{\gamma'}$, respectively, where $\Gamma_a$ and $\Gamma_{\gamma'}$ are the total decay widths of the ALP and the dark photon.
In our benchmark, the relevant total width is approximated by the
dominant two-body partial width in
Eq.~\eqref{eq:gammaPrimedecayrate} or
Eq.~\eqref{eq:ALPdecayrate}, depending on the mass hierarchy.
The percent-level three-body contributions discussed above are neglected.

We show in Fig.~\ref{fig:ctau} a plot of $c\tau_{a/\gamma'}\cdot G^2_{a\gamma\gamma'}$ as a function of $m_{a/\gamma'}$.
The red (blue) curve is for the decay of $\gamma'$ ($a$) with $m_a\ll m_{\gamma'}$ ($m_{\gamma'}\ll m_a$).
The two curves differ by a factor of 3 vertically, as can be inferred from Eq.~\eqref{eq:gammaPrimedecayrate} and Eq.~\eqref{eq:ALPdecayrate}.
It is easy to see that if the value of $G_{a\gamma\gamma'}$ decreases or the mass of the heavier particle is reduced, its lifetime is enhanced,  potentially making it long-lived.

In this work, we propose a search for a single-photon signature arising when the heavier of $\gamma'$ and $a$ decays into the lighter state and a photon.
We require the decay to occur within a three-dimensional lab-frame distance $L_{\rm dec}<L_{\rm prompt}\equiv 1~{\rm cm}$ from the interaction point (IP) and therefore treat the resulting photon as prompt in our analysis.
In the next section, we will discuss our search strategy in detail.

\section{Experiments and simulation}\label{sec:experiments_simulation}

We consider the low-energy $e^+ e^-$ collider experiments BESIII and STCF, focusing on the COM energy $\sqrt{s}=3.097$~GeV at the $J/\psi(1S)$  resonance.
The enhancement in the signal-event yields brought by the resonant production of the $J/\psi$ mesons, compared to the  nonresonant continuum process, is expected to result in excellent sensitivities to the DAP.

Ignoring potential, minor effects of initial-state radiation of the incoming electron and positron beams, we perform  truth-level Monte Carlo (MC) simulations of $J/\psi$ meson decays at rest into an ALP and a dark photon.
The $J/\psi$ mesons considered here are produced through $e^+e^- \to \gamma^* \to J/\psi$ at BESIII and STCF.
In the limit of negligible electron mass, current conservation implies that the longitudinal component of the intermediate virtual photon does not contribute, so that the produced $J/\psi$ mesons are transversely polarized.
Averaging over the two transverse helicity states ($\lambda=\pm1$), the corresponding differential decay rate is
\begin{eqnarray}
    \frac{d\Gamma_T(J/\psi\to a\gamma')}{d\cos\theta_a}   =   \Gamma(J/\psi \to a\gamma') \,  \frac{3}{8} \, (1+\cos^2\theta_a), \label{eq:GammaJpsi_differential}
\end{eqnarray}
where $\theta_a$ denotes the polar angle of the ALP (equivalently for the dark photon, since the two particles are produced back-to-back) with respect to the beam axis, and the subscript $T$ indicates the average over the two transverse polarization states of the $J/\psi$ meson.
This angular distribution is taken into account in the simulation.
We require the heavier of the ALP and dark photon to decay into an SM photon and the lighter particle.
For the  ALP and dark-photon decays, we assume isotropic angular distributions, which is exact for the spin-0 ALP but approximate for the spin-1 dark photon.\footnote{We have checked numerically that the sensitivity results are negligibly affected by the precise choice of the polar-angle distributions.}
We provide in Appendix~\ref{appendix:Jpsi_decay} the analytic differential decay widths of the $J/\psi$ mesons in both signal and irreducible background events.

In the $J/\psi$ rest frame, the photon recoil mass satisfies $m_{\rm rec}^2=m_{J/\psi}^2-2m_{J/\psi}E_\gamma$ and is therefore in one-to-one correspondence with the photon energy, rather than constituting an independent observable.
The recoil system consists of two identical invisible particles, implying $m_{\rm rec}\geq 2m_L$, where $m_L$ denotes the mass of the lighter of $\gamma'$ and $a$.
The recoil-mass distribution is continuous, and both its kinematic endpoints and shape depend on the masses of the two new particles.
With only one observed photon, only the total recoil four-momentum can be reconstructed, while the individual four-momenta of the two invisible particles cannot be determined.

We then impose the following kinematic-acceptance requirements for the final-state photons, in reference to the cuts used by the BESIII collaboration~\cite{BESIII:2017otb}:
$|\cos\theta_\gamma|<0.8$ and $E_{\gamma}>25~\text{MeV}$, where we  consider only the barrel region.
For the photons satisfying these conditions, an efficiency of $99\%$ is applied.
For the STCF experiment, we take the same event-selection cuts as described here.

For each simulated signal event, we apply the probability that the heavier of $a$ and $\gamma'$ decays within a three-dimensional lab-frame distance $L_{\rm dec}<L_{\rm prompt}\equiv 1~{\rm cm}$ from the IP:
\begin{eqnarray}
\epsilon_{\text{prompt}}=1-e^{-L_{\rm prompt}/(\beta \gamma_{\text{L}} c\tau)},  \label{eq:epsilon_prompt}
\end{eqnarray}
where  $\beta=p/E$ is the speed of the decaying particle  with $p$ and $E$ labeling the three-momentum magnitude and energy of the heavier particle, $\gamma_{\text{L}}=E/m$ denotes its Lorentz boost factor with mass $m$, and $c\tau$ is its proper decay length.
For both BESIII and STCF, $L_{\rm prompt}$ is within the beam-pipe vacuum region~\cite{BESIII:2009fln,Ai:2025xop}.
For a photon reaching the barrel electromagnetic calorimeter (EMC) of inner radius $R_{\rm EMC}$, the maximum additional non-pointing angle relative to the corresponding IP-pointing direction is bounded by
$\delta_{\rm np}^{\max}\leq  \arcsin(L_{\rm prompt}/R_{\rm EMC})$.
Using the EMC radii of 94~cm for BESIII~\cite{BESIII:2009fln} and 105~cm for STCF~\cite{Achasov:2023gey}, this additional angle is smaller than $0.61^\circ$ and $0.55^\circ$, respectively.
At the level of our phenomenological treatment, these subdegree deviations are negligible; we therefore apply the same flat $99\%$ efficiency benchmark as for photons originating from the IP.
A dedicated analysis of decays beyond $L_{\rm prompt}$, for which non-pointing-photon information may become relevant, is left for future work.

For the proposed mono-photon search, both reducible and irreducible backgrounds exist.
The reducible backgrounds stem from $e^+ e^-\to \gamma +\cancel{X}$, where the photon $\gamma$ is the only particle detected and $\cancel{X}$ denotes anything undetected by the detector.
Such processes include mainly those with $X$ being a pair of charged fermions, a photon pair, or a single photon.
For the Born two-photon final state, the recoil photon is back-to-back with the tagged photon and therefore lies within the nominal detector acceptance whenever the tagged photon does, in the idealized point-symmetric limit.
Its rejection nevertheless relies on successful reconstruction of the recoil photon.
Geometric gaps, inactive regions, and reconstruction inefficiencies can therefore produce residual two-photon leakage into the mono-photon sample.
The other reducible processes can likewise contribute when the additional final-state particles escape detection.

For reducible configurations in which the undetected particles escape through the nominal forward or backward blind regions, transverse-momentum and energy conservation gives the following upper bound on the energy of the visible photon as a function of its polar angle $\theta_\gamma$~\cite{Liu:2019ogn,Zhang:2019wnz}:
\begin{eqnarray}
    E_{b}^{\text{max}}(\theta_\gamma)=\frac{\sqrt{s}}{1+\sin\theta_\gamma/\sin\theta_{b}},\label{eq:Ebmax}
\end{eqnarray}
where, following Ref.~\cite{Liu:2018jdi}, we have $\sin\theta_b=\sqrt{1-\cos^2\theta_{b}}$ with $|\cos\theta_{b}|=0.95$ corresponding to the boundary of all the sub-detectors at BESIII~\cite{Asner:2008nq}.
We take the same value of $\cos\theta_{b}$ for STCF.
The requirement $E_{\gamma}>E_{b}^{\max}(\theta_\gamma)$ provides a powerful kinematic veto under the idealized assumption of complete detector coverage and perfect veto efficiency down to $\theta_b$.
In a realistic detector, however, geometric gaps, inactive regions, and reconstruction inefficiencies can cause a small fraction of multi-particle events to violate this idealized rejection.
We therefore do not assume that this requirement completely removes the reducible backgrounds.
In the present work, we explicitly estimate only the three-photon contribution and parameterize its migration into the mono-photon signal region by the event-level probability
$\epsilon_{3\gamma}^{\rm leak}$, as detailed in Appendix~\ref{appendix:threegamma_background}.
The Born process $e^+e^-\to2\gamma$, unresolved $2\gamma(+\gamma)$ configurations, radiative-Bhabha backgrounds, and other detector-dependent reducible sources are not included in this benchmark estimate; their migration into the mono-photon signal region requires separate detector-level estimates.
Note that  the cut $E_{\gamma}>E_{b}^{\max}(\theta_\gamma)$ is always stronger than  the $E_\gamma>25$~MeV one in the present study.

The dominant irreducible background stems from $J/\psi\to \gamma \nu\bar{\nu}$ decays, with a branching ratio recently estimated with lattice QCD to be BR$(J/\psi \to \gamma \nu \bar{\nu})=1.04(7)(8)\times 10^{-10}$ in the SM~\cite{Meng:2026ckc}.
We use the lattice result to normalize the total decay rate of $J/\psi\to \gamma \nu\bar{\nu}$, while the kinematic acceptance is evaluated with the leading-order analytic spectrum given below. Uncertainties associated with this spectrum modeling are not included.
Instead of performing a detailed simulation, here we provide  an analytic estimate of the irreducible-background levels at BESIII and STCF.

For BESIII, the total number of resonant $J/\psi$ events collected is determined to be $N^{\text{BESIII}}_{J/\psi}=(10087\pm 44)\times 10^{6}$~\cite{BESIII:2021cxx}.
For STCF, we have $N^{\text{STCF}}_{J/\psi}=3.4\times 10^{12}$ $J/\psi$ mesons per year with an integrated luminosity $\mathcal{L}_{\text{STCF}}=1$~ab$^{-1}$~\cite{Achasov:2023gey}.
Since STCF is planned to operate at $\sqrt{s}=3.097$~GeV for one year, this would lead to non-negligible irreducible-background levels.
We compute analytically the double differential decay rate of $J/\psi\to \gamma \nu\bar{\nu}$ for the transversely polarized $J/\psi$ mesons, and obtain
\begin{eqnarray}
\frac{d^2\Gamma_T(J/\psi\to \gamma\nu\bar{\nu})}{dE_\gamma \, d\cos\theta_\gamma}\propto E_\gamma \Big(  3m_{J/\psi}-4 E_\gamma+\nonumber\\
(4E_\gamma -m_{J/\psi})\cos^2\theta_\gamma \Big), \label{eq:GammaJpsi_differential_bgd}
\end{eqnarray}
with $\cos\theta_\gamma$ ranging between $-1$ and $1$ and $E_\gamma$ between $0$ and $m_{J/\psi}/2$.
The subscript $T$ has the same meaning as that in Eq.~\eqref{eq:GammaJpsi_differential}.

We thus estimate the cut efficiency of $|\cos\theta_\gamma|<0.8$ and $E_\gamma>E_{b}^{\text{max}}(\theta_\gamma)$ with the following formula,
\begin{eqnarray}
\epsilon_{\text{irr}}&=&\frac{\displaystyle \int_{-0.8}^{0.8} d\cos\theta_\gamma\int_{E_{b}^{\text{max}}(\theta_\gamma)}^{m_{J/\psi}/2}dE_\gamma \, \frac{d^2\Gamma_T(J/\psi\to \gamma\nu\bar{\nu})}{dE_\gamma \, d\cos\theta_\gamma}}{\displaystyle \int_{-1}^{1} d\cos\theta_\gamma\int_{0}^{m_{J/\psi}/2}dE_\gamma \, \frac{d^2\Gamma_T(J/\psi\to \gamma\nu\bar{\nu})}{dE_\gamma \,d\cos\theta_\gamma}}\nonumber\\
&\approx& 49.8\%.\label{eq:cut_efficiency_irreducible}
\end{eqnarray}
The quantity $\epsilon_{\rm irr}=49.8\%$ in Eq.~\eqref{eq:cut_efficiency_irreducible} is the kinematic acceptance.
Including the flat single-photon reconstruction efficiency $\epsilon_\gamma \approx 0.99$, the total efficiency is
\begin{equation}
\epsilon_{\rm irr}^{\rm total}
=\epsilon_{\rm irr}\epsilon_\gamma
\approx 49.3\%.
\end{equation}
Using the central value of the SM branching ratio, we therefore obtain
\begin{eqnarray}
N_{\rm irr}^{\rm BESIII}=0.518, N_{\rm irr}^{\rm STCF}=174.5.
\end{eqnarray}

We then derive the signal-event numbers with the following expression:
\begin{eqnarray}
    N_S=N_{J/\psi} \cdot \text{BR}(J/\psi\to a\gamma') \cdot \epsilon, 
\end{eqnarray}
where $\epsilon$ is a selection-efficiency factor, defined as
\begin{eqnarray}
    \epsilon=\frac{1}{N_{\text{MC}}}\sum_{i=1}^{N_{\text{MC}}}\epsilon_{\text{acc.}}^i \cdot \epsilon_{\text{ prompt}}^{i} \cdot \epsilon_{E_{\gamma}}^i,
\end{eqnarray}
where $N_{\text{MC}}$ is the total number of MC simulation events, $\epsilon_{\text{acc.}}^i$ is the acceptance efficiency (including $|\cos\theta_\gamma|<0.8$, $E_\gamma>25$~MeV, and the photon reconstruction efficiency of $99\%$) of the SM photon in the $i^{\text{th}}$ simulated signal event, $\epsilon_{\text{prompt}}^i$ is the prompt-decay probability defined in Eq.~\eqref{eq:epsilon_prompt}, and $\epsilon^i_{E_{\gamma}}=1$ if $E_\gamma^i>E_b^{\text{max}}(\theta_{\gamma}^i)$ and is zero otherwise.

The binary signal-selection factor $\epsilon^i_{E_\gamma}$ should not be confused with $\epsilon_{3\gamma}^{\rm leak}$.
The former is applied to each simulated signal event and implements the requirement $E_\gamma>E_b^{\max}(\theta_\gamma)$, whereas the latter is a net event-level migration probability applied only to the reducible three-photon background.
Since $\epsilon_{3\gamma}^{\rm leak}$ includes reconstruction of the surviving tag photon, no additional factor of $0.99$ is applied to the three-photon yields.

We have also compared the normalized photon-energy and polar-angle distributions of the signal and the irreducible background after the kinematic selections above are imposed.
The distributions substantially overlap over most of the parameter space, with the angular distribution providing little discrimination; some separation in $E_\gamma$ appears only when the heavier new state approaches the kinematic limit, owing to the mass-dependent lower endpoint of the signal spectrum.
We therefore retain an inclusive event-counting analysis for the sensitivity projections, while noting that a dedicated detector-level binned analysis could exploit this high-energy region and sidebands to reduce the dependence on the absolute background normalization.

In the following section, we present the estimated exclusion reach for BESIII and the projected sensitivity for STCF.
For an experiment $X={\rm BESIII,STCF}$, let $N_{3\gamma,X}^{\rm tag}$ denote the number of truth-level resolved three-photon events for which at least one photon satisfies the full tag requirements given in Appendix~\ref{appendix:threegamma_background}, before
detector-induced leakage.
We then define the included benchmark background as
\begin{eqnarray}
B_X=N_{{\rm irr}}^X
+\epsilon_{3\gamma}^{\rm leak}N_{3\gamma,X}^{\rm tag}.
\label{eq:included_background}
\end{eqnarray}

For each leakage benchmark, we adopt the following approximate phenomenological 95\% confidence-level (CL) signal-yield criterion:
\begin{eqnarray}
N_{S,X}^{95}
=\max\left(3,2\sqrt{B_X}\right).
\label{eq:signal_yield_threshold}
\end{eqnarray}
The $2\sqrt{B_X}$ term accounts only for statistical counting fluctuations in the Gaussian approximation, while the three-event floor is retained in the low-background regime.
At each fixed value of $\epsilon_{3\gamma}^{\rm leak}$, $B_X$ is treated as exactly known.
Background-normalization uncertainties associated with ${\cal B}(J/\psi\to\gamma\nu\bar{\nu})$, the three-photon production rates, and the detector-dependent leakage probability are not included
or profiled.
The resulting curves should therefore be interpreted as optimistic phenomenological benchmarks rather than full detector-level expected limits.
The corresponding values are listed in Appendix~\ref{appendix:threegamma_background}.

\section{Numerical results}\label{sec:results}

\begin{figure*}[t]
    \centering
    \includegraphics[width=0.495\textwidth]{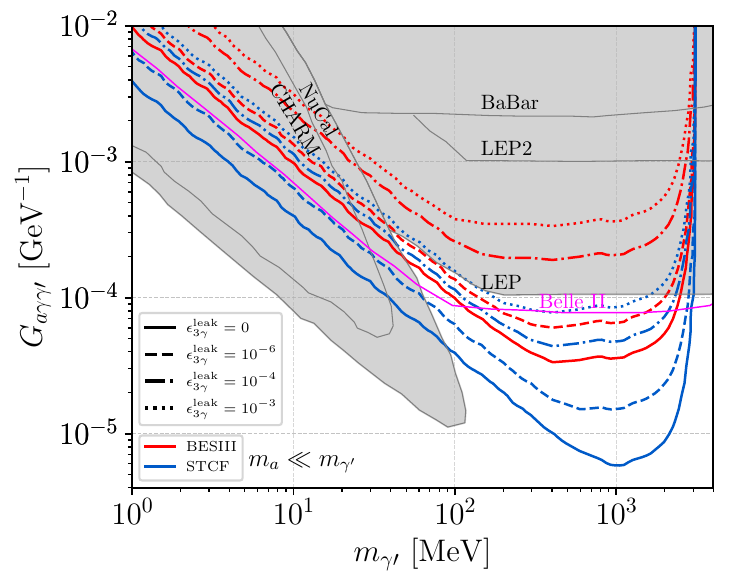}
    \includegraphics[width=0.495\textwidth]{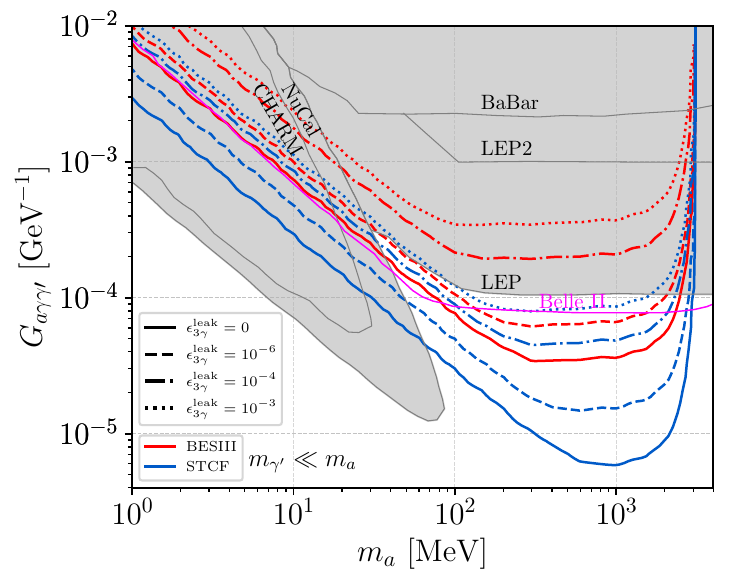}
    \caption{Exclusion reach from the existing BESIII dataset and projected sensitivity of STCF to $G_{a\gamma\gamma'}$ as a function of $m_{\gamma'}$ for $m_a\ll m_{\gamma'}$ (the  left panel) and of $m_a$ for $m_{\gamma'}\ll m_a$ (the right panel). The present upper bounds are displayed in the gray area, obtained from the beam-dump experiments CHARM~\cite{Gninenko:2011uv,Gninenko:2012eq,deNiverville:2018hrc,Jodlowski:2023yne} and NuCal~\cite{Blumlein:1990ay,Blumlein:2011mv,Jodlowski:2023yne}, as well as the $e^+ e^-$ colliders BaBar~\cite{BaBar:2008aby,deNiverville:2018hrc}, LEP~\cite{OPAL:1994kgw,DELPHI:1996drf,Jodlowski:2024ayf}, and LEP2~\cite{DELPHI:2003dlq,Jodlowski:2024ayf}.
    Projected future limits at Belle~II with 50~ab$^{-1}$~\cite{deNiverville:2018hrc,Jodlowski:2024ayf} are also shown.
    All BESIII and STCF curves impose the prompt-decay requirement $L_{\rm dec}<L_{\rm prompt}\equiv 1~{\rm cm}$.
    All the accelerator constraints and projections shown here are based on single-photon signatures.
    Red and blue curves correspond to BESIII and STCF, respectively.
    For each experiment, the solid, dashed, dash-dotted, and dotted curves denote $\epsilon_{3\gamma}^{\rm leak}=0,10^{-6},10^{-4},10^{-3}$, respectively.}
       \label{fig:sensitivities}
\end{figure*}

In Fig.~\ref{fig:sensitivities} we present the estimated exclusion reach  in the coupling $G_{a\gamma\gamma'}$  as a function of $m_{\gamma'}$ for $m_a\ll m_{\gamma'}$ (the left plot) and  of $m_a$ for $m_{\gamma'}\ll m_a$ (the right plot), for the full BESIII dataset already collected (red) and the projected STCF sample (blue).
The results of the BESIII experiment correspond to the $\sim 10^{10}$ $J/\psi$ events that have been collected, and those of STCF are for $N_{J/\psi}^{\text{STCF}}=3.4\times 10^{12}$ expected for an integrated luminosity of $1$~ab$^{-1}$ that can be accumulated with only one year's operation.
Future projections at Belle~II with an integrated luminosity of 50~ab$^{-1}$ are also shown (the magenta curve)~\cite{deNiverville:2018hrc,Jodlowski:2024ayf}.

The two mass hierarchies exhibit similar qualitative behavior.
With increasing mass, the sensitivities of BESIII and STCF are enhanced initially, and then enter a plateau phase,  before deteriorating rapidly near the kinematic endpoint.
At small masses, the heavier particle tends to have a relatively long lifetime, so that its probability of decaying within $L_{\rm prompt}=1~{\rm cm}$ of the interaction point is suppressed,
weakening the sensitivity.
As its mass increases, the lab-frame decay length decreases and the prompt-decay probability approaches unity, producing the plateau behavior shown in the figure.
Finally, for even larger masses approaching $m_{J/\psi}$, the phase-space suppression effects begin to dominate, strongly impairing the sensitivities.

In the zero- and small-leakage benchmarks, BESIII and STCF can probe portal couplings beyond the existing bounds in part of the parameter space.
The precise improvement over current constraints and over the Belle~II projection becomes progressively weaker for larger leakage probabilities and therefore depends on the detector performance.
This advantage is primarily driven by the resonant production and the large samples of $J/\psi$ mesons available at BESIII and STCF, whereas the Belle~II study relies on the nonresonant continuum process $e^+e^-\to a\gamma'$~\cite{deNiverville:2018hrc}.

The zero-leakage curves should be interpreted as idealized best-case sensitivities.
As $\epsilon_{3\gamma}^{\rm leak}$ increases, the large underlying three-photon rate raises the required signal yield and progressively weakens the coupling reach.

In Fig.~\ref{fig:sensitivities} we have assumed either $m_a\ll m_{\gamma'}$ or $m_{\gamma'}\ll m_a$.
If the mass of the lighter particle approaches that of the heavier one, the sensitivities  weaken as a result of the reduced phase space in the decay of the heavier particle.

\section{Conclusions}\label{sec:conclusions}
In this work, we have investigated the sensitivity of the existing BESIII dataset and the future STCF collider to the dark axion portal, via $J/\psi$ decays into an axionlike particle and a dark photon.
The resonant production of the $J/\psi$ mesons at the COM energy of $\sqrt{s}=3.097$~GeV leads to enhanced production yields of the signal events and hence to  strong sensitivity reach.
Concretely, for BESIII we consider $N_{J/\psi}^{\text{BESIII}}=10087\times 10^6$ corresponding to the $J/\psi$ meson events that have been collected, while for STCF we take $N_{J/\psi}^{\text{STCF}}=3.4\times 10^{12}$ corresponding to an integrated luminosity of $1$~ab$^{-1}$.

We have performed Monte Carlo simulations to estimate the signal acceptance and have included both the irreducible $J/\psi\to\gamma\nu\bar{\nu}$ background and a benchmark estimate of the three-photon reducible backgrounds.

The kinematic requirement based on Eq.~\eqref{eq:Ebmax} provides an idealized veto but does not guarantee complete rejection in a realistic detector.
We therefore parameterize migration of taggable three-photon events into the mono-photon signal region by the net event-level probability $\epsilon_{3\gamma}^{\rm leak}$.
For $0\leq\epsilon_{3\gamma}^{\rm leak}\leq10^{-3}$, the included irreducible-plus-three-photon background ranges from approximately $0.52$ to $2.48\times10^4$ events at BESIII and from
$174.5$ to $8.11\times10^6$ events at STCF.
The zero-leakage curves represent the idealized best-case benchmarks, while the finite-leakage curves quantify their dependence on the detector performance.

A reliable determination of the event-level three-photon leakage probability $\epsilon_{3\gamma}^{\rm leak}$, together with separate estimates of the Born-level two-photon, unresolved $2\gamma(+\gamma)$, radiative-Bhabha, and other detector-dependent backgrounds, will require collaboration-level full simulations and data-driven control samples.

We have presented the exclusion reach associated with the existing BESIII dataset, together with the projected sensitivity of STCF, at $95\%$ CL, considering either $m_a\ll m_{\gamma'}$ or $m_{\gamma'}\ll m_a$.
Our numerical results indicate that 
within the zero- and sufficiently small three-photon-leakage benchmarks, the existing BESIII dataset could probe parameter space beyond current bounds, while STCF could substantially extend this reach.
The comparison with Belle~II becomes progressively less favorable as the leakage probability increases.
The favorable reach in the small-leakage regime is primarily driven by resonant production and the large samples of $J/\psi$ mesons available at BESIII and STCF.
Thus, this work illustrates the  potential constraining power of the tau--charm factories on new-physics scenarios predicting new particles arising from rare decays of the $J/\psi$ meson.

\section*{Acknowledgments}
We would like to thank Yongcheng Wu (Nanjing Normal University), Yu Zhang (University of South China), and Xiaorong Zhou (University of Science and Technology of China), for useful discussions.
ZSW and YZ are supported by the National Natural Science Foundation of China under grants No.~12475106 and No.~12505120, and the Fundamental Research Funds for the Central Universities under Grant No.~JZ2025HGTG0252.
DH is supported by the Heze University doctoral fund project No.~010008002039037.


\appendix
\section{Analytic expressions of differential decay rates of $J/\psi$ in both signal and background processes}\label{appendix:Jpsi_decay}

We consider both unpolarized and transversely polarized $J/\psi$ mesons.

For the signal process $J/\psi\to a\gamma'$, the differential decay rate of a transversely polarized $J/\psi$ meson, $\frac{d\Gamma_T(J/\psi\to a\gamma')}{d\cos\theta_a}$, is given by\footnote{As mentioned in Sec.~\ref{sec:intro}, we ignore the $Z$-boson contributions to the signal process in this work since we study low-energy $e^+e^-$ collider experiments BESIII and STCF.}
\begin{eqnarray}
    \frac{d\Gamma_T(J/\psi\to a\gamma')}{d\cos\theta_a} &=& \frac{\alpha f^2_{J/\psi} G^2_{a\gamma\gamma'}}{144m^5_{J/\psi}}\Big(1 + \cos^2\theta_a \Big)\nonumber\\
    &&\lambda^{3/2}(m_{J/\psi}^2,m_a^2,m_{\gamma'}^2)\nonumber\\
    &=&\Gamma(J/\psi \to a\gamma') \,  \frac{3}{8} \, \Big(1+\cos^2\theta_a\Big),\,\quad\quad\label{eqn:differential_decay_rate_signal}
\end{eqnarray}
where we have averaged over the two transverse helicity states by taking a factor $1/2$ and used Eq.~\eqref{eq:GammaJpsi2agammaPrime}.

For an unpolarized $J/\psi$ meson, the differential decay rate $\frac{d\Gamma(J/\psi\to a\gamma')}{d\cos\theta_a}$ is independent of  $\cos\theta_a$ and is simply,
\begin{eqnarray}
    \frac{d\Gamma(J/\psi\to a\gamma')}{d\cos\theta_a} = \frac{\Gamma(J/\psi\to a\gamma')}{2}.
\end{eqnarray}

For the background process $J/\psi\to \gamma \nu \bar{\nu}$, the double differential decay rate $\frac{d^2\Gamma_T(J/\psi\to \gamma\nu\bar{\nu})}{dE_\gamma d\cos\theta_\gamma}$ for one generation of the neutrinos reads,
\begin{eqnarray}
    \frac{d^2\Gamma_T(J/\psi\to \gamma\nu\bar{\nu})}{dE_\gamma d\cos\theta_\gamma}&=&\frac{\alpha G_F^2 |\psi_{J/\psi}(0)|^2 Q_c^2 }{4\pi^2m_{J/\psi}}E_\gamma\Big(3m_{J/\psi}-4 E_\gamma \nonumber\\
    &&+(4E_\gamma -m_{J/\psi})\cos^2\theta_\gamma    \Big),\,\quad\quad\label{eqn:differential_decay_rate_background}
\end{eqnarray}
where $G_F$ is the Fermi constant and the non-perturbative parameter $\psi_{J/\psi}(0)$ is related to the $f_{J/\psi}$ by the following relation which we have derived from Eq.~(13) of Ref.~\cite{Hao:2006nf},
\begin{eqnarray}
    f_{J/\psi}=-2\sqrt{\frac{N_c}{m_{J/\psi}}}\psi_{J/\psi}(0),
\end{eqnarray}
where $N_c=3$ is the color factor of quarks.
Similar to Eq.~\eqref{eqn:differential_decay_rate_signal}, we have averaged over the two transverse helicity states in Eq.~\eqref{eqn:differential_decay_rate_background}.

For an unpolarized $J/\psi$ meson, we find
\begin{eqnarray}
    \frac{d\Gamma(J/\psi\to \gamma\nu\bar{\nu})}{d E_\gamma}=\frac{4\alpha G_F^2 |\psi_{J/\psi}(0)|^2 Q_c^2}{3\pi^2 m_{J/\psi}}E_\gamma(m_{J/\psi}-E_\gamma),\quad\label{eqn:differential_decay_rate_background_unpolarized}
\end{eqnarray}
for one generation of the neutrinos.
Eq.~\eqref{eqn:differential_decay_rate_background_unpolarized} is in agreement with Eq.~(10) of Ref.~\cite{Gao:2014yga}.\footnote{We note, however, in Ref.~\cite{Gao:2014yga}, a factor of $4$ is missing in its Eq.~(7) for the expression of the matrix element of $J/\psi\to \gamma \nu\bar{\nu}$.}

\section{Benchmark estimate of reducible three-photon backgrounds}
\label{appendix:threegamma_background}

We consider three sources of reducible three-photon background to the mono-photon search:
(i) the continuum QED process $e^+e^-\to3\gamma$; (ii) the resonant cascade processes $e^+e^-\to J/\psi\to\gamma P$, followed by $P\to\gamma\gamma$, with $P=\pi^0,\eta,\eta',\eta_c$; and
(iii) the direct resonant process $e^+e^-\to J/\psi\to3\gamma$.
A truth-level three-photon event from any of these sources is classified as taggable if at least one of its photons satisfies
\begin{equation}
|\cos\theta_\gamma|<0.8,\qquad E_\gamma>25~{\rm MeV},\qquad E_\gamma>E_b^{\max}(\theta_\gamma).
\end{equation}
This is an event-level selection: each event is counted at most once.

For the continuum contribution, we generate $e^+e^-\to3\gamma$ at leading order at $\sqrt{s}=3.097~{\rm GeV}$ with \textsc{MadGraph5\_aMC@NLO}~\cite{Alwall:2014hca,Frederix:2018nkq}, using the full SM implementation and the physical electron mass $m_e=0.511\text{~MeV}$.
Each of the three generated photon is required at generation level to satisfy $E_\gamma>25~{\rm MeV}$.
No generator-level angular cut is imposed; in particular, photons with $|\cos\theta_\gamma|>0.95$ are thus retained.
The physical electron mass regulates the initial-state collinear enhancement.
We obtain
\begin{equation}
\sigma_{\rm cont}^{\rm gen}=63.6~{\rm nb},\qquad
A_{\rm cont}=0.1028,
\end{equation}
and therefore
\begin{equation}
\sigma_{\rm cont}^{\rm tag} =\sigma_{\rm cont}^{\rm gen}A_{\rm cont} =6.54~{\rm nb}.
\end{equation}
Here $A_{\rm cont}$ denotes the event-level taggable acceptance under the truth-level selections defined above.

For the continuum contribution only, we further classify the selected events according to $N_{\rm fwd}$, defined as the number of the two non-tag photons satisfying $|\cos\theta_\gamma|\geq0.95$.
If several photons satisfy the full tag selection, the highest-energy one is chosen as the tag. We obtain
\[
\begin{array}{c|ccc}
N_{\rm fwd} & 0 & 1 & 2 \\ \hline
\text{fraction of tagged events}
& 23.38\% & 76.62\% & 0 \\
\sigma_{\rm cont}^{\rm tag}(N_{\rm fwd})\,[{\rm nb}]
& 1.53 & 5.01 & 0
\end{array}
\]
The $N_{\rm fwd}=1$ category matches the topology with one additional photon lying outside the detector boundary and the other evading the detector veto.
For $N_{\rm fwd}=0$, both additional photons must evade the veto.
No $N_{\rm fwd}=2$ event survives the ideal kinematic requirement.

\begin{table}[t]
\begin{ruledtabular}
\begin{tabular}{lcc}
Channel & $A$ & $\text{BR}_{\rm chain}A$ \\ \hline
$\gamma\pi^0$ with $\pi^0\to\gamma\gamma$      & 0.73 & $2.46\times10^{-5}$ \\
$\gamma\eta$ with $\eta\to\gamma\gamma$       & 0.76 & $3.26\times10^{-4}$ \\
$\gamma\eta'$  with $\eta'\to\gamma\gamma$     & 0.80 & $9.80\times10^{-5}$ \\
$\gamma\eta_c$  with $\eta_c\to\gamma\gamma$    & 0.81 & $3.14\times10^{-6}$ \\
direct $3\gamma$   & 0.86 & $9.97\times10^{-6}$
\end{tabular}
\end{ruledtabular}
\caption{Kinematic acceptances and branching-ratio-weighted acceptances for the resonant three-photon backgrounds. The branching ratios are extracted from Ref.~\cite{ParticleDataGroup:2024cfk}.}
\label{tab:resonant_3gamma}
\end{table}

For the resonant cascade modes $J/\psi\to\gamma P$, $P\to\gamma\gamma$, with $P=\pi^0,\eta,\eta',\eta_c$, we use the transverse $J/\psi$ angular distribution proportional to
$1+\cos^2\theta$ and generate the pseudoscalar decay isotropically in the $P$ rest frame.
The direct $J/\psi\to3\gamma$ contribution is modeled with a transversely polarized version of the lowest-order Ore--Powell matrix element~\cite{Ore:1949te,Adkins:1996odo}, analogous to the
ortho-positronium-based signal model used in the BESIII analysis~\cite{BESIII:2012lxx}.
The resulting acceptances and effective branching fractions are summarized in Table~\ref{tab:resonant_3gamma}.

The cascade sum is
\begin{equation}
R_{\gamma P}
=\sum_P {\rm BR}(J/\psi\to\gamma P)
{\rm BR}(P\to\gamma\gamma)A_P
=4.52\times10^{-4},
\end{equation}
while the direct contribution is
\begin{equation}
R_{\rm direct}
={\rm BR}(J/\psi\to3\gamma)A_{3\gamma}
=9.97\times10^{-6}.
\end{equation}
Here $A_P$ and $A_{3\gamma}$ denote the corresponding event-level taggable acceptances for the cascade and direct three-photon samples, respectively.

For experiment $X$ (BESIII/STCF), the number of taggable three-photon events before detector-induced leakage is
\begin{equation}
N_{3\gamma,X}^{\rm tag} ={\cal L}_X\sigma_{\rm cont}^{\rm tag} + N_{J/\psi,X}(R_{\gamma P}+R_{\rm direct}).
\end{equation}
Using ${\cal L}_{\rm BESIII}=3.083~{\rm fb}^{-1}$~\cite{BESIII:2021cxx}, ${\cal L}_{\rm STCF}=1~{\rm ab}^{-1}$~\cite{Achasov:2023gey}, $N_{J/\psi}^{\rm BESIII}=1.0087\times10^{10}$, and
$N_{J/\psi}^{\rm STCF}=3.4\times10^{12}$, we obtain
\begin{equation}
N_{3\gamma,\rm BESIII}^{\rm tag}=2.48\times10^7,
\qquad
N_{3\gamma,\rm STCF}^{\rm tag}=8.11\times10^9.
\end{equation}

We define $\epsilon_{3\gamma}^{\rm leak}$ as the net event-level probability for a taggable three-photon event to be reconstructed in the exactly-one-photon signal region.
It is applied once to each event:
\begin{equation}
N_{3\gamma,X}^{\rm red} =\epsilon_{3\gamma}^{\rm leak} \cdot N_{3\gamma,X}^{\rm tag}.
\end{equation}
Because this probability includes reconstruction of the surviving tag photon, no additional factor of $\epsilon_\gamma=0.99$ is applied to $N_{3\gamma,X}^{\rm red}$.

\begin{table}[t]
\vspace*{6pt}
\setlength{\tabcolsep}{4pt}
\begin{ruledtabular}
\begin{tabular}{cccc}
Experiment
& $\epsilon_{3\gamma}^{\rm leak}$
& $B$
& $N_S^{95}$ \\ \hline
BESIII
& $0$
& $0.518$
& $3$ \\
& $10^{-6}$
& $25.3$
& $10.1$ \\
& $10^{-4}$
& $2.48\times10^3$
& $99.6$ \\
& $10^{-3}$
& $2.48\times10^4$
& $315.0$ \\ \hline
STCF
& $0$
& $174.5$
& $26.4$ \\
& $10^{-6}$
& $8.28\times10^3$
& $182.0$ \\
& $10^{-4}$
& $8.11\times10^5$
& $1800.8$ \\
& $10^{-3}$
& $8.11\times10^6$
& $5694.0$
\end{tabular}
\end{ruledtabular}
\caption{Included benchmark background yields and signal-event thresholds
for the four event-level three-photon leakage benchmarks.}
\label{tab:threegamma_backgrounds}
\end{table}

The signal-event thresholds in Table~\ref{tab:threegamma_backgrounds} are evaluated using
Eq.~\eqref{eq:signal_yield_threshold} and account only for statistical counting fluctuations.
Background-normalization uncertainties are not included.

The continuum and resonant contributions are added incoherently, and interference effects are neglected.
The continuum result corresponds to a generator-level three-photon sample defined by the requirement $E_\gamma>25~{\rm MeV}$ for each photon.
The Born process $e^+e^-\to2\gamma$ is not included.
Configurations containing an additional photon below this generation-level energy cutoff, referred to here as unresolved $2\gamma(+\gamma)$ backgrounds, are likewise not included.
Their detector-induced migration is not described by $\epsilon_{3\gamma}^{\rm leak}$ and would require a separate detector-level leakage estimate.
We also do not include radiative-Bhabha backgrounds, higher-order QED corrections, detector-response or background-normalization systematic uncertainties, or correlations among inefficiencies.
Accordingly, the values of $\epsilon_{3\gamma}^{\rm leak}$ should be regarded as detector-performance benchmarks rather than predictions.
Their determination requires full detector simulations and data-driven control samples.

\FloatBarrier
\bibliography{refs}

\end{document}